\def\be{\begin{equation}}
\def\ee{\end{equation}}
\def\ba{\begin{array}}
\def\ea{\end{array}}

\documentclass[aps,amsmath,amssymb,amsfonts,showpacs]{revtex4}
\usepackage{graphicx}
\usepackage{epstopdf}
\def\qed{\leavevmode\unskip\penalty9999 \hbox{}\nobreak\hfill
     \quad\hbox{\leavevmode  \hbox to.77778em{%
               \hfil\vrule   \vbox to.675em%
               {\hrule width.6em\vfil\hrule}\vrule\hfil}}
     \par\vskip3pt}

\newtheorem{theorem}{Theorem}
\newtheorem{corollary}{Corollary}
\newtheorem{lemma}{Lemma}

\begin{document}

\title{Monogamy Properties of Qubit Systems}

\author{Xue-Na Zhu$^{1}$}
\author{Shao-Ming Fei$^{2,3}$}

\affiliation{$^1$School of Mathematics and Statistics Science, Ludong University, Yantai 264025, China\\
$^2$School of Mathematical Sciences, Capital Normal University,
Beijing 100048, China\\
$^3$Max-Planck-Institute for Mathematics in the Sciences, 04103 Leipzig, Germany}

\bigskip

\begin{abstract}

We investigate monogamy relations related to quantum entanglement for $n-$qubit quantum systems.
General monogamy inequalities are presented to the $\beta$th $(\beta\in(0,2))$ power
of concurrence, negativity and the convex-roof extended negativity, as well as the $\beta$th $(\beta\in(0,\sqrt{2}))$ power of entanglement of formation. These monogamy relations
are complementary to the existing ones with different regions of parameter $\beta$.
In additions, new monogamy relations are also derived which include the existing ones as special cases.

\end{abstract}

\maketitle

\section{INTRODUCTION}
Quantum entanglement \cite{F,K,H,J,C} lies at the heart of quantum
information processing and quantum computation\cite{ma}. Accordingly its quantification has drawn much attention
in the last decade.
As one of the fundamental differences between quantum entanglement and classical
correlations, a key property of entanglement is that a quantum system entangled with one of other
systems limits its entanglement with the remaining systems.
The monogamy relations give rise to the structures of entanglement distribution in
multipartite systems. Monogamy is also an essential feature
allowing for security in quantum key distribution \cite{k3}.

For a tripartite system $A$, $B$ and $C$,
the monogamy of an entanglement measure $\varepsilon$ implies that \cite{022309},
the entanglement between $A$ and $BC$ satisfies
$\varepsilon_{A|BC}\geq\varepsilon_{AB}+\varepsilon_{AC}$.
Such monogamy relations are not always satisfied by any entanglement measures.
It has been shown that the squared concurrence $C^2$ \cite{PRA80044301,C2}
and the squared entanglement of formation $E^2$ \cite{PRLB,PRA61052306}
do satisfy such monogamy relations. In
Ref.\cite{zhuxuena} it has been shown that the general monogamy inequalities are satisfied by the $\alpha(\alpha\geq2)$th power
of concurrence $C^\alpha$ and the $\alpha(\alpha\geq\sqrt{2})$th power
of entanglement of formation $E^{\alpha}$ for $n-$qubit mixed states.
Another useful entanglement
measure is the negativity\cite{gv}, a quantitative version of Peres's criterion for separability.
The authors in Ref.\cite{n1} studied
the monogamy property of the $\alpha$th power of negativity $N^{\alpha}~(\alpha\geq2)$ and
discussed tighter $\alpha$th $(\alpha\geq2)$ power
of the convex-roof extended negativity (CREN) $\tilde{N}^{\alpha}$.
In Ref.\cite{jin} tighter monogamy inequalities for concurrence, entanglement of formation and CREN has been investigated for $\alpha\geq2$.

However, it is not clear for the monogamy properties of the $\alpha$th $(0<\alpha<2)$ power of concurrence, negativity and  CREN,
and the $\alpha$th $(0<\alpha<\sqrt{2})$ power of entanglement of formation.
In this paper, we study the general monogamy inequalities of $C^{\beta}$, $N^{\beta}$, $\tilde{N}^{\beta}$ and $E^{\beta}$ for $\beta\in[0,M]$, where $M$ is any real number greater than zero.

\section{MONOGAMY PROPERTY OF CONCURRENCE}

For a bipartite pure state $|\psi\rangle_{AB}$,
the concurrence is given by \cite{s7,s8,af},
\begin{equation}\label{CON}
C(|\psi\rangle_{AB})=\sqrt{2[1-Tr(\rho^2_A)]},
\end{equation}
where $\rho_A$ is reduced density matrix by tracing over the subsystem $B$,
$\rho_{A}=Tr_{B}(|\psi\rangle_{AB}\langle\psi|)$.
The concurrence is extended to mixed states $\rho=\sum_{i}p_{i}|\psi _{i}\rangle \langle \psi _{i}|$,
$p_{i}\geq 0$, $\sum_{i}p_{i}=1$, by the convex roof construction,
\begin{equation}\label{CONC}
C(\rho_{AB})=\min_{\{p_i,|\psi_i\rangle\}} \sum_i p_i C(|\psi_i\rangle).
\end{equation}

For $n-$qubit quantum states, the concurrence satisfies \cite{zhuxuena}
\begin{eqnarray}\label{a1}
C^{\alpha}_{A|B_1B_2...B_{n-1}}\geq C^{\alpha}_{AB_1}+...+C^{\alpha}_{AB_{n-1}},
\end{eqnarray}
for $\alpha\geq 2$,
where $C_{A|B_1B_2...B_{n-1}}$ is the concurrence of  $\rho$ under bipartite
partition $A|B_1B_2...B_{n-1}$, and $C_{AB_i}$, $i=1,2...,n-1$, is the
concurrence of the mixed states $\rho_{AB_i}=Tr_{B_1B_2...B_{i-1}B_{i+1}...B_{n-1}}(\rho)$.
For $C_{AB_i}\not=0$, $i=1,...,n-1$, the concurrence satisfies
\begin{equation}\label{a2}
C^{\alpha}_{A|B_{1}...B_{n-1}}
<C^{\alpha}_{AB_{1}}+...+C^{\alpha}_{AB_{n-1}},
\end{equation}
for $\alpha\leq0$. Further, in Ref. \cite{jin} tighter monogamy inequalities than (\ref{a1}) are derived for the $\alpha$th $(\alpha\geq2)$ power of concurrence.

\begin{lemma}\label{La1}
For real numbers $x\in[0,1]$  and $t\geq1$, we have $(1+t)^x\geq1+(2^x-1)t^x.$
\end{lemma}
{\sf [Proof]}~Let $g_{x}(t)=\frac{(1+t)^x-1}{t^x}$ with $x\in[0,1]$ and $t\in[1,+\infty)$.
Since $\frac{d g_{x}(t)}{d t}=xt^{-(x+1)}[1-(1+t)^{x-1}]\geq0$, we obtain that $g_{x}(t)$ is an
increasing function of $t$. Hence, $g_{x}(t)\geq g_{x}(1)$, i.e, $(1+t)^x\geq1+(2^x-1)t^x.$

\begin{theorem}\label{TH1}
For any $2\otimes2\otimes 2^{n-2}$ tripartite mixed state:

(1) if $C_{AB}\leq C_{AC}$, the concurrence satisfies
\begin{equation}\label{ca1}
C^{\beta}_{A|BC}\geq
C^{\beta}_{AB}+(2^{\frac{\beta}{\alpha}}-1)C^{\beta}_{AC},
\end{equation}
where $0\leq\beta\leq\alpha$ and $\alpha\geq 2$.

(2) if $C_{AB}\geq C_{AC}$, the concurrence satisfies
\begin{equation}\label{ca2}
C^{\beta}_{A|BC}\geq
(2^{\frac{\beta}{\alpha}}-1)C^{\beta}_{AB}+C^{\beta}_{AC},
\end{equation}
where
$0\leq\beta\leq\alpha$ and $\alpha\geq 2$.
\end{theorem}

{\sf [Proof]}~ For arbitrary $2\otimes2\otimes2^{n-2}$ tripartite
state $\rho_{ABC}$, one has \cite{zhuxuena},
 \begin{equation}\label{c12}\nonumber
 C^{\alpha}_{A|BC}\geq C^{\alpha}_{AB}+C^{\alpha}_{AC}.
 \end{equation}
If $\max\{C_{AB},C_{AC}\}=0$, i.e., $C_{AB}=C_{AC}=0$, obviously we have
the inequalities (\ref{ca1}) or (\ref{ca2});
If $\min\{C_{AB},C_{AC}\}=0$, obviously $C^{\beta}_{A|BC}\geq\max\{C^{\beta}_{AB},C^{\beta}_{AC}\}\geq (2^{\frac{\beta}{\alpha}}-1)\max\{C^{\beta}_{AB},C^{\beta}_{AC}\}$ with $0\leq\beta\leq\alpha$,
we also have the inequalities (\ref{ca1}) or (\ref{ca2}).

If $\max\{C_{AB},C_{AC}\} >0$ and $\min\{C_{AB},C_{AC}\}\not=0$, assuming $0< C_{AB}\leq C_{AC}$, we have
\begin{eqnarray}\nonumber
C^{\alpha x}_{A|BC}
&\geq& (C^{\alpha}_{AB}+C^{\alpha}_{AC})^{x}\\\nonumber
&=&C^{\alpha x}_{AB}
\Big(1+\frac{C^{\alpha}_{AC}}{C^{\alpha}_{AB}}\Big)^{x}\\\nonumber
&\geq&C^{\alpha x}_{AB}
\left(1+(2^x-1)\Big(\frac{C^{\alpha}_{AC}}{C^{\alpha}_{AB}}\Big)^{x}\right)\\\nonumber
&=&C^{\alpha x}_{AB}+(2^x-1)C^{\alpha x}_{AC},
\end{eqnarray}
where the second inequality is due to the inequality $(1+t)^x\geq1+(2^x-1)t^x$ for $0\leq x\leq1$ and $t=\frac{C^{\alpha}_{AC}}{C^{\alpha}_{AB}}\geq1$.
Denote $\alpha x=\beta$. Then $\beta\in [0,\alpha]$ since $x\in[0,1]$ and one gets the inequality (\ref{ca1}).
If $C_{AB}\geq C_{AC}$, similar proof gives the inequality (\ref{ca2}).

One can see that Theorem \ref{TH1} reduces to the monogamy inequality (\ref{a1})
if $\beta=\alpha\geq 2$. In particular, if we take $\beta=1$,
we have $C_{A|BC}\geq\min\{C_{AB},C_{AC}\}+(2^{\frac{1}{\alpha}}-1)
\max\{C_{AB},C_{AC}\}$ for $\alpha\geq2$. And the tighter relation is $C_{A|BC}\geq\min\{C_{AB},C_{AC}\}+(\sqrt{2}-1)
\max\{C_{AB},C_{AC}\}$.

{\it Example 1.} Let us
consider the three-qubit case. Any three-qubit state $|\psi\rangle$ can be written in
the generalized Schmidt decomposition \cite{zhuxuena,gx,X},
\begin{equation}\label{pure1}
|\psi\rangle=\lambda_0|000\rangle+
\lambda_1e^{i\varphi}|100\rangle
+\lambda_2|101\rangle
+\lambda_3|110\rangle+
\lambda_4|111\rangle,
\end{equation}
where $\lambda_i\geq0$, $i=0,...,4$, and $\sum_{i=0}^{4}\lambda_i^2=1$.
From  Eq.(\ref{CON}) and Eq.(\ref{CONC}), we have
$C_{A|BC}=2\lambda_0\sqrt{\lambda^2_2+\lambda^2_3+\lambda^2_4},$
$C_{AB}=2\lambda_0\lambda_2,$ and $C_{AC}=2\lambda_0\lambda_3.$
Without loss of generality, we set $\lambda_0=\cos\theta_0,$
$\lambda_1= \sin\theta_0\cos\theta_1,$ $\lambda_2=
\sin\theta_0\sin\theta_1\cos\theta_2,$ $\lambda_3=
\sin\theta_0\sin\theta_1\sin\theta_2\cos\theta_3, $ and $\lambda_4=
\sin\theta_0\sin\theta_1\sin\theta_2\sin\theta_3$,
$\theta_i\in[0,\frac{\pi}{2}]$.
Assume $\lambda_3\geq\lambda_2$, i.e $C_{AC}\geq C_{AB}:$

(a) if $\theta_2=\frac{\pi}{2}$, we have
\begin{eqnarray*}
C^{\beta}_{A|BC}-C^{\beta}_{AB}-(2^{\frac{\beta}{\alpha}}-1)C^{\beta}_{AC}
&=&(2\lambda_0)^{\beta}
\Big[(\lambda^2_2+\lambda^2_3+\lambda^2_4)^{\frac{\beta}{2}}-\lambda^{\beta}_2
-(2^{\frac{\beta}{\alpha}}-1)\lambda^{\beta}_3\Big]\\
&=&(2\lambda_0)^{\beta}\sin^{\beta}\theta_0\sin^{\beta}\theta_1
\Big[
1-(2^{\frac{\beta}{\alpha}}-1)\cos^{\beta}\theta_3
\Big]\\
&\geq&(2\lambda_0)^{\beta}\sin^{\beta}\theta_0\sin^{\beta}\theta_1(2-2^{\frac{\beta}{\alpha}})\\
&\geq&0,
\end{eqnarray*}
where $0\leq\beta\leq\alpha$, $\alpha\geq2$ and the first inequality is due to $cos\theta_3\leq1$;

(b) if $\theta_2\not=\frac{\pi}{2}$,
we denote $t_1=\frac{\sin^{\alpha}\theta_2\cos^{\alpha}\theta_3}{cos^{\alpha}\theta_2}$.
We have
\begin{eqnarray*}
C^{\beta}_{A|BC}-C^{\beta}_{AB}-(2^{\frac{\beta}{\alpha}}-1)C^{\beta}_{AC}
&=&(2\lambda_0)^{\beta}
\Big[(\lambda^2_2+\lambda^2_3+\lambda^2_4)^{\frac{\beta}{2}}-\lambda^{\beta}_2
-(2^{\frac{\beta}{\alpha}}-1)\lambda^{\beta}_3\Big]\\
&=&(2\lambda_0)^{\beta}\sin^{\beta}\theta_0\sin^{\beta}\theta_1
\Big[
1-\cos^{\beta}\theta_2-(2^{\frac{\beta}{\alpha}}-1)\sin^{\beta}\theta_2\cos^{\beta}\theta_3
\Big]\\
&=&(2\lambda_0)^{\beta}\sin^{\beta}\theta_0\sin^{\beta}\theta_1
\left[1-cos^{\beta}\theta_2\left(1+(2^{\frac{\beta}{\alpha}}-1)t^{\frac{\beta}{\alpha}}_1
\right)\right]\\
&\geq&(2\lambda_0)^{\beta}\sin^{\beta}\theta_0\sin^{\beta}\theta_1
\left[1-cos^{\beta}\theta_2(1+t_1)^{{\frac{\beta}{\alpha}}}\right]\\
&=&(2\lambda_0)^{\beta}\sin^{\beta}\theta_0\sin^{\beta}\theta_1
\Big[1-(cos^{\alpha}\theta_2+sin^{\alpha}\theta_2cos^{\alpha}\theta_3)^{\beta}\Big]\\
&\geq&0,
\end{eqnarray*}
where $0\leq\beta\leq\alpha$ and $\alpha\geq2$. The first inequality is due to Lemma \ref{La1} with $0\leq x=\frac{\beta}{\alpha}\leq1$ and the second inequality is due to
$cos^{\alpha}\theta_2+sin^{\alpha}\theta_2cos^{\alpha}\theta_3\leq 1$ for $\alpha\geq2$.

Therefore, for this case we have $C^{\beta}_{A|BC}\geq C^{\beta}_{AB}+(2^{\frac{\beta}{\alpha}}-1)C^{\beta}_{AC}$ for $0\leq\beta\leq\alpha$ and $\alpha\geq2$.
For the case $\lambda_3\leq\lambda_2$, i.e., $C_{AB}\geq C_{AC}$,
similarly one obtains that $C^{\beta}_{A|BC}\geq (2^{\frac{\beta}{\alpha}}-1)C^{\beta}_{AB}+C^{\beta}_{AC}$ with $0\leq\beta\leq\alpha$ and $\alpha\geq2$.

By using the Theorem \ref{TH1} repeatedly, we have the following theorem for multipartite qubit systems.

\begin{theorem}\label{TH2}
For any $n$-qubit quantum state $\rho$ such that
$C_{AB_i}\leq C_{A|B_{i+1}...B_{n-1}}$ for  $i=1,...,m,$  and
$C_{AB_j}\geq C_{A|B_{j+1}...B_{n-1}}$ for $j=m+1,...,n-2$,
$\forall1\leq m\leq n-3$, $n\geq4,$ we have
\begin{eqnarray}\label{thc2}
C^{\beta}(\rho_{A|B_1B_2...B_{n-1}})
&\geq&
\sum_{i=1}^{m}(2^{\frac{\beta}{\alpha}}-1)^{i-1}C^{\beta}(\rho_{AB_i})\\\nonumber
&+&(2^{\frac{\beta}{\alpha}}-1)^{m+1}\sum_{i=m+1}^{n-2}C^{\beta}(\rho_{AB_i})
+(2^{\frac{\beta}{\alpha}}-1)^{m}C^{\beta}(\rho_{AB_{n-1}}),
\end{eqnarray}
where $0\leq\beta\leq\alpha$ and $\alpha\geq 2$.
\end{theorem}

{\sf [Proof]}~For convenience, we denote $r=2^{\frac{\beta}{\alpha}}-1$.
For any $2\otimes2\otimes2\otimes...\otimes2$ quantum states $\rho_{AB_1...B_{n-1}},$
we have
\begin{eqnarray*}\nonumber
&&C^{\beta}_{A|B_1B_2...B_{n-1}}(\rho)\\
&\geq & C^{\beta}_{AB_1}+rC^{\beta}_{A|B_2...B_{n-1}}\\\nonumber
&\geq& C^{\beta}_{A|B_1}+rC^{\beta}_{A|B_2}
+r^2C^{\beta}_{A|B_3...B_{n-2}}\\\nonumber
&\geq&...\\\nonumber
&\geq&\sum_{i=1}^{m}r^{i-1}C^{\beta}_{AB_i}
+r^{m}C^{\beta}_{A|B_{m+1}...B_{n-1}}\\\nonumber
&\geq&\sum_{i=1}^{m}r^{i-1}C^{\beta}_{AB_i}
+r^{m}\left[rC^{\beta}_{AB_{m+1}}
+C^{\beta}_{A|B_{m+2}...B_{n-1}}\right]\\\nonumber
&\geq&...\\\nonumber
&\geq&\sum_{i=1}^{m}r^{i-1}C^{\beta}_{AB_i}
+r^{m+1}\sum_{i=m+1}^{n-2}C^{\beta}_{AB_i}
+r^{m}C^{\beta}_{AB_{n-1}},
\end{eqnarray*}
where  the first four inequalities are due to  $C_{AB_i}\leq C_{A|B_{i+1}...B_{n-1}}$ $(i=1,...,m)$ and the inequality (\ref{ca1}),
the last three inequalities are due to  $C_{AB_j}\geq C_{A|B_{j+1}...B_{n-1}}$ $(j=m+1,...,n-2)$ and the inequality (\ref{ca2}).

For an $n$-qubit quantum state $\rho_{AB_1...B_{n-1}}$, in Ref.\cite{zhuxuena} it has been shown that the $\beta$th concurrence $C^{\beta}$ $(0<\beta<2)$ does not satisfy monogamy inequalities like $C^{\beta}(|\psi\rangle_{A|B_1B_2...B_{N-1}})\geq\sum_{i=1}^{n-1}C^{\beta}(\rho_{AB_i})$.
Theorem (\ref{TH2}) first time gives the monogamy inequality satisfied by
the $\beta-$th concurrence $C^{\beta}$ for the case of $(0<\beta<2)$,
a problem that was not solved in Refs.\cite{zhuxuena,jin}.
Specifically, if $\beta=1$ and $\alpha=2$, we get the monogamy relation
satisfied by the concurrence $C$:
\begin{eqnarray*}
C(\rho_{A|B_1B_2...B_{N-1}})
\geq
\sum_{i=1}^{m}(\sqrt{2}-1)^{i-1}C(\rho_{AB_i})
+(\sqrt{2}-1)^{m+1}\sum_{i=m+1}^{n-2}C(\rho_{AB_i})
+(\sqrt{2}-1)^{m}C(\rho_{AB_{n-1}}).
\end{eqnarray*}

{\it Example 2.}
Let us consider the pure state $|\psi\rangle$ (\ref{pure1}) in the Example 1.
Set $\lambda_0=\frac{\sqrt{2}}{2}$, $\lambda_1=\frac{1}{2}$,
$\lambda_2=\frac{1}{4}$, $\lambda_3=\frac{3\sqrt{3}}{20}$
and $\lambda_4=\frac{\sqrt{3}}{5}$. We have $C_{A|BC}=\frac{\sqrt{2}}{2}\approx0.707$ and
$C_{AB}+C_{AC}
=\frac{5\sqrt{2}+3\sqrt{6}}{20}\approx0.721.$
One can see that $C_{A|BC}<C_{AB}+C_{AC}$.
Denoting $u(\beta,\alpha)=C^{\beta}_{A|BC}-C^{\beta}_{AB}-(2^{\frac{^{\beta}}{\alpha}}-1)C^{\beta}_{AC}
=(\frac{\sqrt{2}}{2})^{\beta}-(\frac{\sqrt{2}}{4})^{\beta}-(2^{\frac{1}{\alpha}}-1)(\frac{3\sqrt{6}}{20})^{\beta}$ with $0\leq\beta\leq\alpha$ and $\alpha\geq2,$
we have $u(1,\alpha)\geq0.201$ for all $\alpha\geq2$.
Furthermore, our result shows that $u(\beta,\alpha)\geq0$
for all $0\leq\beta\leq2$ and $\alpha\geq2$, see Fig. 1.
\begin{figure}
  \centering
  \includegraphics[width=7cm]{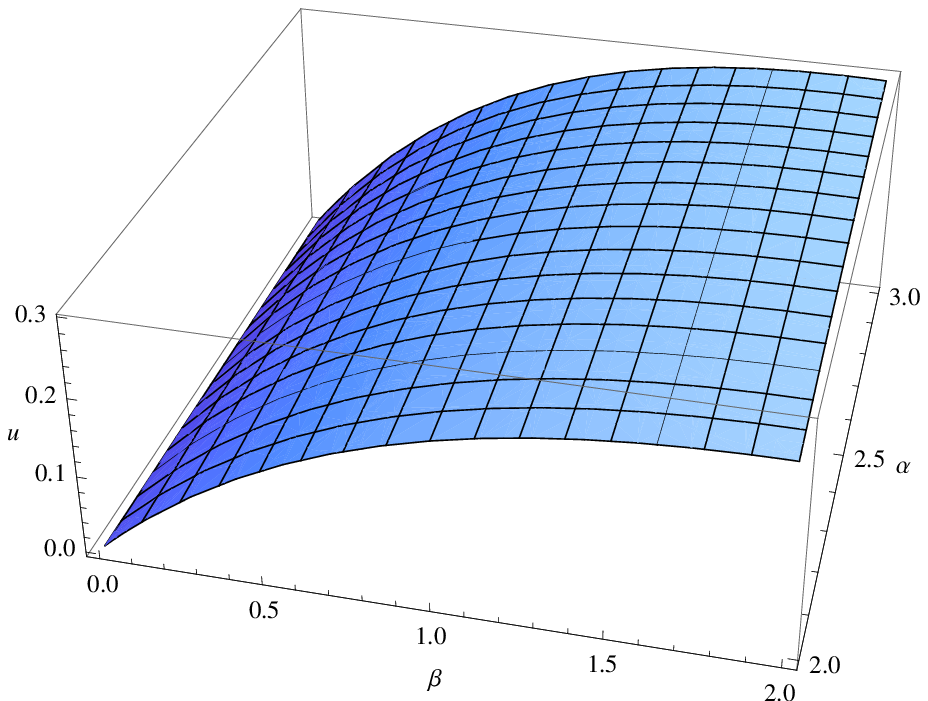}\\
\caption{$u(\beta,\alpha)$ for $0\leq\beta\leq2$ and $\alpha\geq2$.}
\label{fig1}
\end{figure}

\section{MONOGAMY INEQUALITY FOR NEGATIVITY}

Given a bipartite state $\rho_{AB}$, the negativity is defined by \cite{D. P. DiVincenzo}
\begin{eqnarray*}
N(\rho_{AB})=\frac{||\rho^{T_A}_{AB}||-1}{2},
\end{eqnarray*}
where $\rho^{T_A}_{AB}$
is the partially transposed matrix of $\rho_{AB}$ with respect to the
subsystem $A$, $||X||=Tr\sqrt{XX^{\dag}}$ denotes the trace norm of $X$.
For the convenience of discussion,
we use the following definition of negativity:
\begin{eqnarray*}
N(\rho_{AB})=||\rho^{T_A}_{AB}||-1.
\end{eqnarray*}

It has been shown that for any $n$-qubit pure state
$|\psi\rangle_{A|B_1...B_{n-1}}$, the negativity satisfies the monogamy inequality
holds for $\alpha\geq2$ \cite{n1}:
\begin{eqnarray*}
N^{\alpha}_{A|B_1...B_{n-1}}(|\psi\rangle)\geq  N^{\alpha}_{AB_1}+...+N^{\alpha}_{AB_{n-1}},
\end{eqnarray*}
and the polygamy inequality for $\alpha\leq0:$
\begin{eqnarray*}
N^{\alpha}_{A|B_1...B_{n-1}}(|\psi\rangle)< N^{\alpha}_{AB_1}+...+N^{\alpha}_{AB_{n-1}}.
\end{eqnarray*}
Here $N_{A|B_1...B_{n-1}}(|\psi\rangle)$ is the negativity of $|\psi\rangle$
under bipartite partition $A|B_1...B_{n-1}$, and $N_{AB_i}$ is the negativity of the quantum state  $\rho_{AB_i}=Tr_{B_1...B_{i-1}B_{i+1}...B_{n-1}}(|\psi\rangle)$.
In the following we study the monogamy property of the $\beta$th
power of negativity $N^{\beta}$ for $\beta\in(0,2)$.

\begin{theorem}\label{n1}
For any $n$-qubit quantum pure state $|\psi\rangle$ such that
$C_{AB_i}\leq C_{A|B_{i+1}...B_{n-1}}$ for $i=1,...,m$, and
$C_{AB_j}\geq C_{A|B_{j+1}...B_{n-1}}$ for $j=m+1,...,n-2$,
$\forall1\leq m\leq n-3$ and $n\geq4,$ we have
\begin{eqnarray}
N^{\beta}(|\psi\rangle_{A|B_1B_2...B_{N-1}})&\geq&
\sum_{i=1}^{m}(2^{\frac{\beta}{\alpha}}-1)^{i-1}N^{\beta}(\rho_{AB_i})\\\nonumber
&+&(2^{\frac{\beta}{\alpha}}-1)^{m+1}\sum_{i=m+1}^{n-2}N^{\beta}(\rho_{AB_i})
+(2^{\frac{\beta}{\alpha}}-1)^{m}N^{\beta}(\rho_{AB_{n-1}}),
\end{eqnarray}
where $0\leq\beta\leq\alpha$ and $\alpha\geq 2$.
\end{theorem}

Theorem \ref{n1} can be seen by using (\ref{thc2}) in theorem \ref{TH2},
and noting that $C(|\psi\rangle_{A|BC}=N(|\psi\rangle_{A|BC})$
for $2\otimes t\otimes s$ $(t\geq2,~s\geq2)$ systems
and $N(\rho_{AB})\leq C(\rho_{AB})$ for $2\otimes m$ systems.

Given a bipartite state $\rho_{AB}$, the CREN is defined as the convex roof extended
negativity of pure states \cite{n1,Lee}
\begin{eqnarray*}
\tilde{N}(\rho_{AB})=\min_{\{p_i,|\psi_i\rangle\}}\sum_ip_iN(|\psi_i\rangle),
\end{eqnarray*}
with the infimum taking over all possible decompositions of
$\rho_{AB}$ in a mixture of pure states,
$\rho_{AB}=\sum_ip_i|\psi_i\rangle \langle \psi_i|$, $p_i\geq0$, $\sum_ip_i=1$.

For a mixed state $\rho_{ABC}$ in $2\otimes 2\otimes 2^{n-2}$
systems, the following monogamy inequality holds \cite{n1},
$\tilde{N}^{\alpha}_{A|BC}(\rho)\geq  \tilde{N}^{\alpha}_{AB}+\tilde{N}^{\alpha}_{AC}$
for $\alpha\geq2$,
and the following polygamy inequality holds,
$\tilde{N}^{\alpha}_{A|BC}(\rho)< \tilde{N}^{\alpha}_{AB}+\tilde{N}^{\alpha}_{AC}$
for $\alpha\leq0$.
For multiqubit mixed states $\rho_{AB_1...B_{n-1}}$, one has
the following monogamy inequality for the $\alpha$th power of CREN
for $\alpha\geq2$\cite{n1}:
\begin{eqnarray*}
\tilde{N}^{\alpha}_{A|B_1...B_{n-1}}(\rho)\geq  \tilde{N}^{\alpha}_{AB_1}+...+\tilde{N}^{\alpha}_{AB_{n-1}},
\end{eqnarray*}
 and the following polygamy inequality for $\alpha\leq0$:
\begin{eqnarray*}
\tilde{N}^{\alpha}_{A|B_1...B_{n-1}}(\rho)< \tilde{N}^{\alpha}_{AB_1}+...+\tilde{N}^{\alpha}_{AB_1...B_{n-1}},
\end{eqnarray*}
where $\tilde{N}_{A|B_1...B_{n-1}}(\rho)$ is the negativity of $\rho$ under
bipartite partition $A|B_1...B_{n-1}$, and $\tilde{N}_{AB_i}$ is
the negativity of the quantum state  $\rho_{AB_i}=Tr_{B_1...B_{i-1}B_{i+1}...B_{n-1}}(\rho)$.

With a similar consideration to concurrence, we obtain the following result.
\begin{corollary}\label{n2}
For any $2\otimes2\otimes 2^{n-2}$  mixed state $\rho_{ABC}$, and
$0\leq\beta\leq\alpha$, $\alpha\geq 2$:

(1) if $\tilde{N}_{AB}\leq \tilde{N}_{AC}$, the CREN  satisfies
\begin{equation}\label{nn1}
\tilde{N}^{\beta}_{A|BC}\geq
\tilde{N}^{\beta}_{AB}+(2^{\frac{\beta}{\alpha}}-1)\tilde{N}^{\beta}_{AC};
\end{equation}

(2) if $\tilde{N}_{AB}\geq \tilde{N}_{AC}$, the CREN satisfies
\begin{equation}\label{nn2}
\tilde{N}^{\beta}_{A|BC}\geq
(2^{\frac{\beta}{\alpha}}-1)\tilde{N}^{\beta}_{AB}+\tilde{N}^{\beta}_{AC}.
\end{equation}
\end{corollary}

\begin{corollary}\label{n3}
For any $n$-qubit quantum state $\rho_{AB_1...B_{n-1}}$ such that
$\tilde{N}_{AB_i}\leq \tilde{N}_{A|B_{i+1}...B_{n-1}}$ $(i=1,...,m)$ and
$\tilde{N}_{AB_j}\geq \tilde{N}_{A|B_{j+1}...B_{n-1}}$ $(j=m+1,...,n-2)$,
$\forall1\leq m\leq n-3$, $n\geq4,$ we have
\begin{eqnarray*}\label{nnn2}
\tilde{N}^{\beta}(\rho_{A|B_1B_2...B_{N-1}})\geq
\sum_{i=1}^{m}(2^{\frac{\beta}{\alpha}}-1)^{i-1}\tilde{N}^{\beta}(\rho_{AB_i})
+(2^{\frac{\beta}{\alpha}}-1)^{m+1}\sum_{i=m+1}^{n-2}\tilde{N}^{\beta}(\rho_{AB_i})
+(2^{\frac{\beta}{\alpha}}-1)^{m}\tilde{N}^{\beta}(\rho_{AB_{n-1}}),
\end{eqnarray*}
where $0\leq\beta\leq\alpha$ and $\alpha\geq 2$.
\end{corollary}

\section{MONOGAMY INEQUALITY FOR EoF}

The entanglement of formation (EoF) \cite{C. H. Bennett,D. P. DiVincenzo} is
a well-defined and important measure of quantum entanglement for bipartite systems.
Let $H_A$ and $H_B$ be $m$- and $n$-dimensional $(m\leq n)$ vector spaces, respectively.
The EoF of a pure state $|\psi\rangle\in H_A\otimes H_B$ is defined by
$E(|\psi\rangle)=S(\rho_A)$,
where $\rho_A=Tr_{B}(|\psi\rangle\langle\psi|)$
and $S(\rho)=Tr(\rho\log_2\rho)$. For a bipartite
mixed state $\rho_{AB}\in H_A\otimes H_B$, the entanglement of formation is given by
\begin{eqnarray*}
E(\rho_{AB})=\min_{\{p_i,|\psi_i\rangle\}}\sum_ip_iE(|\psi_i\rangle),
\end{eqnarray*}
with the infimum taking over all possible decompositions of
$\rho_{AB}$ in a mixture of pure states
$\rho_{AB}=\sum_ip_i|\psi_i\rangle \langle \psi_i|$, where
$p_i\geq0$ and $\sum_ip_i=1$.

Denote $f(x)=H\left(\frac{1+\sqrt{1-x}}{2}\right),$
where $H(x)=-x\log_2(x)-(1-x)\log_2(1-x).$
One has \cite{jin}
\begin{equation}\label{ef}
E(\rho_{AB})\geq f(C^2_{AB}).
\end{equation}

\begin{lemma}\label{La2}
If $0\leq x\leq y\leq1$, we have
\begin{equation}\label{f}
f^{\beta}(x^2+y^2)\geq f^{\beta}(x^2)+(2^{\frac{\beta}{\alpha}}-1)f^{\beta}(y^2),
\end{equation}
where $f^{\beta}(x^2+y^2)=(f(x^2+y^2))^{\beta}$, $ 0\leq\beta\leq\alpha$ and $\alpha\geq\sqrt{2}.$
\end{lemma}

{\sf [Proof]}~
Since $0\leq x\leq y\leq1$ and $f(x)$ is a  monotonically increasing  function for $0\leq x\leq1$, one has $f(x^2)\leq f(y^2)$ and $f^{\alpha}(x^2+y^2)
\geq f^{\alpha}(x^2)+f^{\alpha}(y^2)$ for $\alpha\geq\sqrt{2}$ \cite{zhuxuena}.
Let $z\in[0,1]$.

If $x=0$, i.e., $f(x^2)=0$, we have
\begin{eqnarray*}\nonumber
f^{\alpha z}(x^2+y^2)
&\geq&(f^{\alpha}(x^2)+f^{\alpha}(y^2))^{z}\\
&=&(f^{\alpha}(y^2))^{z}\\
&\geq&(2^z-1)f^{\alpha z}(y^2)).
\end{eqnarray*}

If $x\not=0$, i.e., $f(x^2)\not=0$, we have
\begin{eqnarray*}\nonumber
f^{\alpha z}(x^2+y^2)
&\geq&(f^{\alpha}(x^2)+f^{\alpha}(y^2))^{z}\\
&=&f^{\alpha z}(x^2)\left(1+\frac{f^{\alpha}(y^2)}{f^{\alpha}(x^2)}\right)^{z}\\
&\geq&f^{\alpha z}(x^2)+(2^z-1)f^{\alpha z}(y^2)),
\end{eqnarray*}
where the last inequality is obtained by using lemma \ref{La1}.
The Lemma \ref{La2} is proved by setting $\alpha z=\beta$.

It has been shown that the entanglement of formation does not satisfy
monogamy inequality such as $E_{AB}+E_{AC}\leq E_{A|BC}$ \cite{PRA61052306}.
In \cite{zhuxuena} the authors showed that
$E^{\alpha}(\rho_{A|B_1B_2...B_{n-1}})\geq \sum_{i=1}^{n-1}E^{\alpha}(\rho_{AB_i})$
for $\alpha\geq\sqrt{2}$,
and $E^{\alpha}(\rho_{A|B_1B_2...B_{n-1}})\leq \sum_{i=1}^{n-1}E^{\alpha}(\rho_{AB_i})$
for $\alpha\leq0$. In
Ref. \cite{jin} tighter monogamy relation for $E^{\alpha}(\alpha\geq\sqrt{2})$
has been derived for $n$-qubit states.

In fact, applying the same approach to the theorems \ref{TH1} and \ref{TH2},
we can prove the following results generally:

\begin{theorem}\label{TH3}
For any $2\otimes2\otimes 2$ mixed state $\rho\in H_A\otimes H_{B}\otimes H_{C}$,
and $0\leq\beta\leq\alpha$, $\alpha\geq \sqrt{2}$.

(1) If $C_{AB}\leq C_{AC}$, we have
\begin{equation}\label{ce1}
E^{\beta}_{A|BC}\geq
E^{\beta}_{AB}+(2^{\frac{\beta}{\alpha}}-1)E^{\beta}_{AC};
\end{equation}

(2) If $C_{AB}\geq C_{AC}$, we have
\begin{equation}\label{ce2}
E^{\beta}_{A|BC}\geq
(2^{\frac{\beta}{\alpha}}-1)E^{\beta}_{AB}+C^{\beta}_{AC}.
\end{equation}
\end{theorem}

{\sf [Proof]}~Let $\alpha\geq\sqrt{2}$ and $\beta\in[0,\alpha]$.
If $C_{AB}\leq C_{AC}$, we have
 \begin{eqnarray*}\nonumber
 E^{\beta}_{A|BC}&\geq&f^{\beta}(C^2_{A|BC})\\
 &\geq& f^{\beta}(C^2_{AB}+C^2_{AC})\\
  &\geq&f^{\beta}(C^2_{AB})+(2^{\frac{\beta}{\alpha}}-1)f^{\beta}(C^2_{AC})\\
  &=&E_{AB}^{\beta}+(2^{\frac{\beta}{\alpha}}-1)E_{_{AC}}^{\beta},
 \end{eqnarray*}
where the first inequality is due to the inequality (\ref{ef}), the second inequality is obtained from the inequality $C_{A|BC}^2\geq C_{AB}^2+C_{AC}^2$, the third inequality holds because of Lemma \ref{La2} and the last equality is obtained from $E(\rho)=f(C^2(\rho))$ for two qubit states.
The result for the case 2 can be similarly proved.

{\it Example 3.}
Consider the $W$ state, $|W\rangle=\frac{1}{\sqrt{3}}(|100\rangle+|010\rangle+|00 1\rangle)$.
We have $E_{A|BC}=0.918296$, $E_{AB}=E_{AC}=0.550048$. Therefore, $E_{A|BC}<E_{AB}+E_{AC}$.
It is easily verified that $E_{A|BC}>0.897968=\max_{\alpha\geq\sqrt{2}}\left(E_{AB}+(2^{\frac{1}{\alpha}}-1)E_{AC}\right)$.
Denote $u(\beta,\alpha)=E^{\beta}_{A|BC}-E^{\beta}_{AB}-(2^{\frac{\beta}{\alpha}}-1)E^{\beta}_{AC}
=0.918296^{\beta}-2^{\frac{\beta}{\alpha}}\times0.550048^{\beta}$. For $0\leq\beta\leq1$ and $\alpha\geq\sqrt{2}$, we have $u(\beta,\alpha)\geq0$, see Fig.2.
\begin{figure}
  \centering
  \includegraphics[width=7cm]{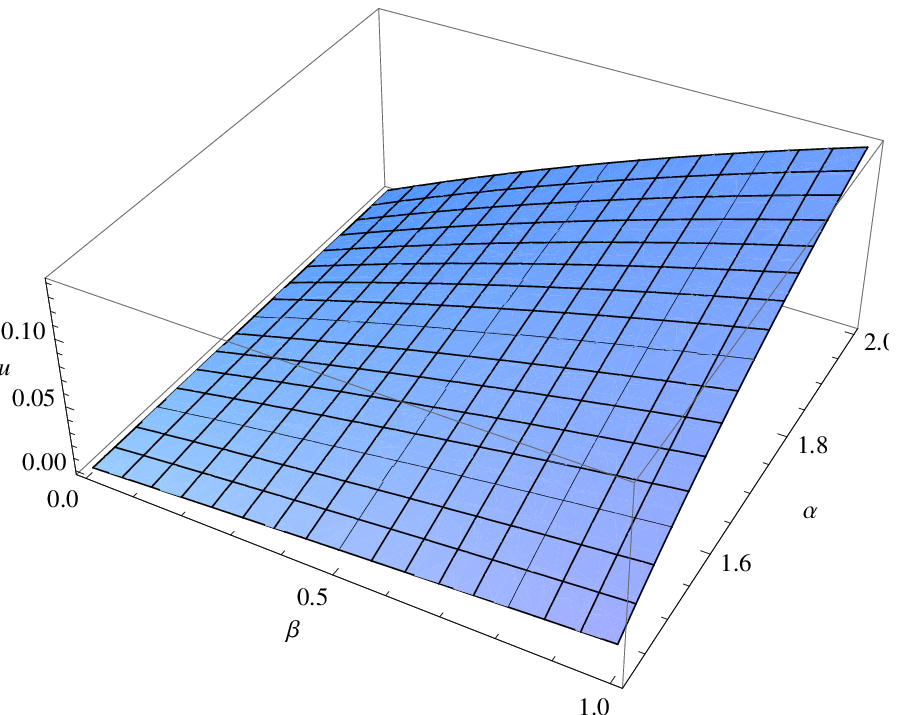}\\
\caption{$u(\beta,\alpha)$ for $0\leq\beta\leq1$ and $\alpha\geq\sqrt{2}$.}
\label{fig2}
\end{figure}

For $n-$qubit quantum states, we have follow theorem.
\begin{theorem}\label{TH4}
For any $n-$qubit mixed state $\rho_{AB_1...B_{n-1}}$ such that
$C_{AB_i}\leq C_{A|B_{i+1}...B_{n-1}}$ $(i=1,...,m)$  and
$C_{AB_j}\geq C_{A|B_{j+1}...B_{n-1}}$ $(j=m+1,...,n-2)$,
$\forall1\leq m\leq n-3$ and $n\geq4,$ we have
\begin{eqnarray*}
E_{A|B_1B_2...B_{N-1}}^{\beta}(\rho)\geq
\sum_{i=1}^{m}(2^{\frac{\beta}{\alpha}}-1)^{i-1}E^{\beta}(\rho_{AB_i})
+(2^{\frac{\beta}{\alpha}}-1)^{m+1}\sum_{i=m+1}^{n-2}E^{\beta}(\rho_{AB_i})
+(2^{\frac{\beta}{\alpha}}-1)^{m}E^{\beta}(\rho_{AB_{n-1}}),
\end{eqnarray*}
where $0\leq\beta\leq\alpha$ and $\alpha\geq\sqrt{2}$, $E_{A|B_1B_2...B_{n-1}}$
is the entanglement of formation of $\rho$ under bipartite
partition $A|B_1B_2...B_{n-1}$, and $E_{AB_i}$, $i=1,2...,n-1$, is the
entanglement of formation of the mixed state $\rho_{AB_i}
=Tr_{B_1B_2...B_{i-1}B_{i+1}...B_{n-1}}(\rho)$.
\end{theorem}

{\sf [Proof]}~ Denote $k=2^{\frac{\beta}{\alpha}}-1$.
For $\alpha\geq\sqrt{2}$ and $\beta\in[0,\alpha]$, we have
\begin{eqnarray}\label{e1}
E^{\beta}_{A|B_1B_2...B_{N-1}}
&\geq& f^{\beta}(C^2_{A|B_1B_2...B_{N-1}})\\\nonumber
&\geq& f^{\beta}(C^2_{AB_1}+C^2_{A|B_2...B_{n-1}})\\\nonumber
&\geq& f^{\beta}(C^2_{AB_1})
+kf^{\beta}(C^2_{A|B_2...B_{n-1}})\\\nonumber
&\geq&...\\\nonumber
&\geq&\displaystyle\sum_{i=1}^{m}k^{i-1}f^{\beta}(C^2_{AB_i})
+k^{m}f^{\beta}(C^2_{A|B_{m+1}...B_{n-1}})\\\nonumber
&=&\displaystyle\sum_{i=1}^{m}k^{i-1}E^{\beta}(\rho_{AB_i})+k^{m}f^{\beta}(C^2_{A|B_m+1...B_{n-1}}),
\end{eqnarray}
where the first inequality is due to (\ref{ef}), the third to the fifth inequalities are due to
$C_{A|B_i}\leq C_{A|B_{i+1}...B_{n-1}}$ $(i=1,...,m)$ and Lemma \ref{La2}. Moreover,
\begin{eqnarray}\label{e2}
f^{\beta}(C^2_{A|B_m+1...B_{n-1}})
&\geq& f^{\beta }(C^2_{AB_{m+1}}+C^2_{A|B_{m+2}...B_{n-1}})\\\nonumber
&\geq& kf^{\beta}(C^2_{A|B_{m+1}})+f^{\beta }(C^2_{A|B_{m+2}...B_{n-1}})\\\nonumber
&\geq&...\\\nonumber
&\geq& k\sum_{i=m+1}^{n-2}f^{\beta}(C^2_{A|B_{i}})+f^{\beta }(C^2_{AB_{n-1}})\\\nonumber
&=&k\sum_{i=m+1}^{n-2}E^{\beta}(\rho_{AB_i})+E^{\beta}(\rho_{AB_{n-1}}),
\end{eqnarray}
where the second to  the fourth inequalities are due to
$C_{AB_i}\geq C_{A|B_{i+1}...B_{n-1}}(i=m+1,...,n-2)$ and Lemma \ref{La2}.

Combining (\ref{e1})and (\ref{e2}) we obtain
the theorem \ref{TH4}.

Theorem \ref{TH4} gives the monogamy relations satisfied by the $\beta$th
($0\leq\beta\leq\alpha$, $\alpha\geq\sqrt{2}$) power of EoF for $n$-qubit states,
which is a problem remained unsolved in Ref.\cite{zhuxuena,jin} for $\beta\in(0,\sqrt{2})$.
If we take $\beta=\alpha\geq\sqrt{2},$ Theorem \ref{TH4} reduces to the result in Ref.\cite{zhuxuena}.
In addition if we take $\beta=1$ and $\alpha=\sqrt{2}$ for theorem \ref{TH4}, we have
\begin{eqnarray*}
E(|\psi\rangle_{A|B_1B_2...B_{n-1}})\geq
\sum_{i=1}^{m}(2^{\frac{1}{\sqrt{2}}}-1)^{i-1}E(\rho_{AB_i})
+(2^{\frac{1}{\sqrt{2}}}-1)^{m+1}\sum_{i=m+1}^{n-2}E(\rho_{AB_i})
+(2^{\frac{1}{\sqrt{2}}}-1)^{m}E(\rho_{AB_{n-1}}),
\end{eqnarray*}
which gives first time the tight monogamy inequality satisfied by the entanglement of formation itself.

\section{CONCLUSION}

Entanglement monogamy is a fundamental property of multipartite
entangled states. We have investigated the monogamy relations
related to the concurrence, the negativity, CREN and the
entanglement of formation for general $n$-qubit states.
We have derived the monogamy inequalities satisfied by
$C^{\beta}$, $N^{\beta}$, $\tilde{N}^{\beta}$ for $\beta\in(0,2)$,
and $E^{\beta}$ for $\beta\in(0,\sqrt{2})$ for $n$-qubit states.
These monogamy relations are complementary to the existing ones
with different regions of parameter $\beta$.
Our new monogamy relations also include the existing ones as special cases.
Our approach may be used to study further monogamy properties
related to other quantum entanglement measures such as Tsallis-$q$ entanglement and to
quantum correlations such as quantum discord.

\bigskip
\noindent{\bf Acknowledgments}\, \, This work is supported by NSFC under numbers 11675113, 11605083, and Beijing Municipal Commission of Education (KM201810011009).

\end{document}